\documentclass{elsart}
\usepackage{graphicx}

\def\Journal#1#2#3#4{{#1} {\bf #2} (#4) #3}

\def\NPB{Nucl. Phys. B}
\def\NPBOLD{Nucl. Phys.}

\def\PLB{{Phys. Lett.} B}

\def\PLBOLD{Phys. Lett.}
\def\PRL{Phys. Rev. Lett.}

\def\PRD{Phys. Rev. D}

\def\JHEP{JHEP}

\def\mapleq{\mathbin{\lower.3ex\hbox{$\buildrel<\over{\smash{\scriptstyle\sim}\vphantom{_x}}$}}}
\begin{document}
\begin{frontmatter}
\title{LMA MSW Solution in the Minimal $SU(3)_L \times U(1)_N$ Gauge Model
\thanksref{talk}
}
\thanks[talk]{Talk given by T.K. at the International Workshop NuFACT'01, Tsukuba, Japan (May 2001).}
\author{Teruyuki Kitabayashi$^a$, Masaki Yasu\`{e}$^{b,c}$}
\address{$^a$
         {\small \sl Accelerator Engineering Center,\\
         Mitsubishi Electric System \& Service Engineering Co.Ltd.,\\
		 2-8-8 Umezono, Tsukuba, Ibaraki 305-0045, Japan.
		 }
		}
\address{$^b$
        {\small \sl Department of Natural Science,\\
        School of Marine Science and Technology, Tokai University,\\
        3-20-1 Orido, Shimizu, Shizuoka 424-8610, Japan
		}
		}
\address{$^c$
        {\small \sl Department of Physics, Tokai University,\\
        1117 Kitakaname, Hiratsuka, Kanagawa 259-1291, Japan.
		}
		}
\begin{abstract}
The minimal $SU(3)_L \times U(1)_N$ gauge model for $(\nu,\ell,\ell^C)$ equipped with the (approximate) $L_e-L_\mu-L_\tau$ symmetry and a discrete $Z_4$ symmetry is found to provide radiative neutrino masses compatible with the LMA MSW solution.
\end{abstract}
\end{frontmatter}

The neutrino oscillation phenomena imply the existence of massive neutrinos \cite{MassiveNu}. Non-zero Majorana neutrino masses are known to arise in models with lepton number violating interactions and $SU(3)_L \times U(1)_N$ models naturally have such interactions \cite{SU3U1}, which are linked to generate charged lepton masses. With this in mind, we have constructed the simplest $SU(3)_L \times U(1)_N$ model with for $(\nu,\ell,\ell^C)$ giving rise to the currently most favorable LMA solution \cite{KitaMinimal}.

We use the following three guidelines to construct our model. (1) the model accounts for $\Delta m_{atm}^2 \gg \Delta m_\odot^2$ \cite{Observation}, which will be realized by bimaximal neutrino mixings based on the (approximate) $L^\prime \equiv L_e-L_\mu-L_\tau$ symmetry \cite{Lprime}; (2) the model respects the LMA solution, $\Delta m_{atm}^2/\Delta m_\odot^2 = 10^{-2} - 10^{-1}$ \cite{Observation}; and (3) the model provides neutrino masses as radiative effects \cite{one-loop}.

The particle content is given by, for $(U(1)_N, L^\prime)$: 
(a) triplet leptons, 
$\psi^1=(\nu^1,\ell^1,\ell^{c1})_L^T:(0,1), 
\psi^{i=2,3}=(\nu^i,\ell^i,\ell^{ci})_L^T:(0,-1) 
$; 
(b) triplet scalars,
$ \eta=(\eta^0,\eta^-,\eta^+)^T    :(0,0),
  \rho=(\rho^+,\rho^0,\rho^{++})^T :(1,0),
  \chi=(\chi^-,\chi^{--},\chi^0)^T :(-1,0)
$ with $\langle \eta^0,\rho^0,\chi^0\rangle$ ($\equiv$ $v_{\eta,\rho,\chi}$) 
and
(c) antisextet scalars, $S^{(0)}: (1,0)$ for $\psi^1\psi^{2,3}$, $S^{(+)}: (1,2)$ for $\psi^{2,3}\psi^{2,3}$. If there is only one antisextet, because the antisextet yields both mass terms for $\ell$ and $\nu$, we will encounter the similar flavor mass-structure for $\nu$ and $\ell$, which, however, contradicts with the phenomenology. To avoid this 'similarity', we introduce two antisextets of $S^{(0,+)\alpha\beta}$ ($\alpha,\beta$=1,2,3) and study their vacuum alignment to yield, for neutrinos, $\langle S^{(0)11}\rangle$ ($\equiv v^{(0)}_\nu$) but without $\langle S^{(+)11}\rangle$ ($\equiv v^{(+)}_\nu$) and, for charged leptons, $\langle S^{(0,+)23}\rangle$ ($\equiv v^{(0,+)}_\ell$) \cite{KitaMinimal}.

To realize the vacuum alignment, we use the $L^\prime$ symmetry with a $Z_4$ symmetry, where $Z_4(\psi^1,\eta, S^{(+)}) $=+; $Z_4(\psi^{2,3},S^{(0)})$=$-$; $Z_4(\rho,\chi)=i$. The interaction, $\eta\eta S^{(+)c} S^{(+)c}$, which is $Z_4$ (and $L$)-symmetric for $S^{(+)11}$, does not disturb $v^{(+)}_\nu$=0 and breaks the $L^\prime$ conservation but preserves a discrete symmetry based on exp(i$L^\prime\pi$/2) ($\equiv$ $Z_{L^\prime}$) since $L^\prime$ is broken by $\vert L^\prime \vert$ = 4. Then, $Z_4$ and $Z_{L^\prime}$ symmetries collaborate to ensure the vacuum alignment and all possible $L$-violating interactions that can generate $v^{(+)}_\nu$ turn out to be forbidden.

\begin{figure}[t]
  \begin{center}
    \includegraphics*[30mm,245mm][190mm,280mm]{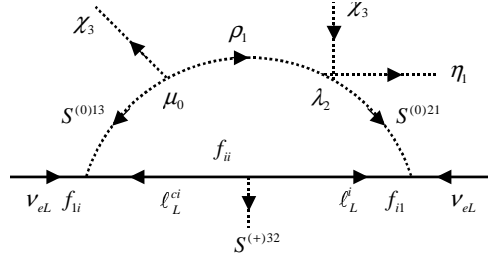}
    \caption{\label{Fig:oneloop}
  One-loop diagram for $\delta m_{11}^{rad}$, where $m_{\ell^i}=f_{ii}v_\ell^{(+)}$ .}
  \end{center}
\end{figure}

The Yukawa interactions consist of 
$f_{1i} \overline{\left( \psi^1 \right)^C} \psi^i S^{(0)}$ and 
$(1/2)f_{ij} \overline{\left( \psi^i \right)^C} \psi^j$ $S^{(+)}$ for $i,j$ = 2,3 (without $\psi\psi\eta$ forbidden by $Z_4$) and Higgs potentials are given by trivial self-Hermitian terms and by non self-Hermitian terms composed of $\mu_0 \rho S^{(0)}\chi$ and $\lambda_1 \eta\eta S^{(0)C}S^{(0)C}$ generating $v^{(0)}_\ell \neq$0 and $\lambda_2 \eta S^{(0)C} \rho^C \chi^C$ generating $v^{(0)}_\nu \neq$0, both of which conserve $L^\prime$, and of $\lambda_b \eta\eta S^{(+)C}S^{(+)C}$ generating $v^{(+)}_\ell \neq$0, which violates $L^\prime$.  These Higgs interactions support our vacuum alignment.

We obtain neutrino and charged lepton mass matrices as follows: 
\begin{equation}
M_\nu = \left( 
    \begin{array}{ccc}
    \delta m_{11}^{rad} & \epsilon\delta m_{12}^\ell & \epsilon\delta m_{13}^\ell \\
    \epsilon\delta m_{12}^\ell   & 0                 & 0        \\
    \epsilon\delta m_{13}^\ell   & 0                 & 0
    \end{array}
\right), \quad
M_\ell = \left( 
    \begin{array}{ccc}
    0 & \delta m_{12}^\ell & \delta m_{13}^\ell \\
    \delta m_{12}^\ell   & m_{22}^\ell                 & m_{23}^\ell        \\
    \delta m_{13}^\ell   & m_{23}^\ell                 & m_{33}^\ell
    \end{array}
\right)
\label{Eq:Mnu}
\end{equation}
where $\delta m_{1i}^\ell$ $(m_{ij}^\ell)$ = $f_{1i}v^{(0)}_\ell$ ($f_{ij}v^{(+)}_\ell$) ($i,j$=2,3), $\epsilon$ = $v^{(0)}_\nu /v^{(0)}_\ell$, $\delta m_{11}^{rad}$ is a radiative mass as in Fig.\ref{Fig:oneloop}: $\delta m_{11}^{rad} \sim -2\lambda_2 f_{13}^2 m_\tau \mu_0 v_\chi^2 /v_{weak}^3 $ ($v_{weak}\equiv$ 174 GeV).  The hierarchical parameterization of $m_{ij}^\ell$ is used for $m_\mu \ll m_\tau$ and $\delta m_{1i}^\ell \ll m_{ij}^\ell$ for $m_e\ll m_{\mu,\tau}$ \cite{HierachicalLeptons} and the bimaximal structure relies upon $\delta m_{12}^\ell\approx \delta m_{13}^\ell$.

Using the phenomenologically acceptable parameter set of ($\mu_0=m_\chi=v_\chi$, $v_{\eta,\rho,S^{(0,+)}}=m_{\rho, S^{(0,+)}}=v_{weak}/2$) and an estimate of $v^{(0)}_\nu\sim -\lambda_2 v_\eta v_\rho/v_\chi$ \cite{Lambda} with $\lambda_2\sim 10^{-7}$ for $m_\nu$ $\sim$ $5\times 10^{-3}$ eV$^2$, we finally obtain 
$\Delta m_\odot^2/\Delta m_{atm}^2$ $\sim$ $2\delta m_{11}^{rad}/m_\nu$ $\sim$ $(10^{-2}-10^{-1})$ for $v_\chi \sim 3.5-6.0$ TeV, which are consistent with the  $SU(3)_L$ breaking scale.  This model is, thus, relevant to yield the LMA solution. \footnote{The ratio $\Delta m_\odot^2/\Delta m_{atm}^2$ is also induced by the diagonalization effects of $M_\ell$ estimated to be $\mapleq (m_e/m_\mu)^{3/2}\approx 3.4\times 10^{-4}$, which can be safely neglected \cite{KitaMinimal}.}

We have constructed the minimal $SU(3)_L \times U(1)_N$ model with the leptons placed in $(\nu,\ell,\ell^C)$. The approximate bimaximal mixings are based on the $L^\prime$ symmetry and the specific emphasis is laid on the fact that the similarity of our neutrino mass matrix to the charged lepton mass matrix for the (1,2) and (1,3)-entries in the $\epsilon$ terms of Eq.(\ref{Eq:Mnu}). We found sufficient power of symmetries that forbids neutrino mass terms in the ($i,j$)-entries ($i,j$ = 2,3) while these terms are allowed for charged leptons. The observed mass hierarchy $\Delta m_{atm}^2 \gg \Delta m_\odot^2$ is explained by the dynamical one between tree and one-loop effects and the LMA solution shows up for $v_\chi\sim$a few TeV. 

The work of M.Y. is supported by the Grant-in-Aid for Scientific Research No 12047223 from the Ministry of Education, Science, Sports and Culture, Japan.




\end{document}